# Dynamic response of exchange bias in graphene nanoribbons

S. Narayana Jammalamadaka[a, b*], S. S. Rao[c, d, e*], J. Vanacken[a], V. V. Moshchalkov[a]

*[a]INPAC – Institute for Nanoscale Physics and Chemistry, KU Leuven, Celestijnenlaan 200D, B–3001 Leuven, Belgium*

*[b]Department of Physics, Indian Institute of Technology Hyderabad, Ordnance Factory Estate Yeddumailaram, Andhra Pradesh, India, 502205.*

*[c]INPAC – Institute for Nanoscale Physics and Chemistry, Semiconductor Physics Laboratory,  KU Leuven, Celestijnenlaan 200D, B–3001 Leuven, Belgium*

*[d]Materials Science Division, Army Research Office, Research Triangle Park, North Carolina 27709, USA*

*[e]Department of Material Science and Engineering, North Carolina State University, Raleigh, North Carolina 27695, USA*

Wei Lu[1] and J. M. Tour[1,2,3]

*[1]Department of Chemistry, [2]Department of Mechanical Engineering and Materials Science, [3]Smalley Institute for Nanoscale Science and Technology, Rice University, MS-222, 6100  Main Street, Houston, Texas 77005, USA.*

## Abstract

The dynamics of magnetic hysteresis, including the training effect and the field sweep rate dependence of the exchange bias, is experimentally investigated in exchange-coupled potassium split graphene nanoribbons (GNRs). We find that, at low field sweep rate, the pronounced absolute training effect is present over a large number of cycles. This is reflected in a gradual decrease of the exchange bias with the sequential field cycling. However, at high field sweep rate above 0.5 T/min, the training effect is not prominent. With the increase in field sweep rate, the average value of exchange bias field grows and is found to follow power–law behavior. The response of the exchange bias field to the field sweep rate variation is linked to the difference in the time it takes to perform a hysteresis loop measurement compared with the relaxation time of the anti-ferromagnetically aligned spins. The present results may broaden our current understanding of magnetism of GNRs and would be helpful in establishing the GNRs-based spintronic devices.

**Keywords:** Dynamic magnetic properties, exchange bias, training effect, field sweep rate and graphene nanoribbons

*These authors contributed equally to this work*

*Corresponding authors:* surya@iith.ac.in , ssingam@ncsu.edu

Ever since the discovery of graphene, the demand for carbon based materials has been rising as they would have broad applications specifically in the area of information storage, information processing, high speed communication and low power consumption[1 - 5]. On the other hand, the unconventional magnetism of the carbon based materials has been of great interest in perspective of spin–based applications as graphene would offer a possibility of tuning its spin–transport properties by means of various applied conditions[6, 7]. The discovery[8] of weak ferromagnetism in polymerized $C_{60}$ has invoked a special attention to investigate the magnetic properties of carbon-based materials. Graphene is an allotrope of carbon and irradiation of graphene with ions or electrons has led to the appearance of magnetism emerging as a result of removal of carbon atoms from the graphene layer, which gives quasi localized states at the Fermi level[9 - 14].

Density functional theory (*DFT*) of single atom vacancies in graphene has disclosed that the accounted magnetic moments are equal to 1.12–1.53 $\mu_B$ per vacancy depending on the defect concentration. Several groups[10, 13] also argued that if the defects are in the same sub-lattice, there would be ferromagnetic (FM) coupling among the spins. Efforts[11, 15] pertinent to the effect of hydrogen adsorption on graphene uncovered the existence of magnetic moments on neighboring carbon atoms and localized spin polarized states around adsorptive hydrogen.

Graphene nanoribbons (GNRs) are thin layers of graphene, characterized by magnetic zigzag edges, which are apparently absent in graphene[16 – 18] in its pristine form. GNRs with zigzag edges have narrow-band edge states at the Fermi energy (*FE*), implying possible magnetization at the edges. First principle calculations[16, 19] for such GNRs with zigzag edges have demonstrated the existence of long range magnetic ordering among the edge states. Theoretically, it has been believed[16, 19] that such long range magnetic ordering is *FM* if there is a coupling between the spins that residing on the same edges, however, this would be anti–FM (*AFM)* if the coupling occurs between opposite edges. Such theoretical predictions for *GNRs* had not been paralleled by experimental observations until our report[20] of the comprehensive experimental findings on the magnetism of GNRs.

In our earlier work[20], we reported on extensive experimental insights pertinent to the magnetic property comparisons of graphene nanoribbons (*GNRs*) (prepared by potassium splitting of carbon nanotubes[21]) with those of chemically converted graphene nanoribbons (*CCGNRs*) (oxidativly unzipped carbon nanotubes that were then chemically reduced[22]). We have shown that *GNRs* exhibit room temperature *FM*-like properties, on the other hand, the *CCGNRs* reveal low temperature (< 20 K) *FM*-like properties and such features are found to be absent in the latter ribbons at room temperature. GNRs are shown to exhibit negative exchange bias (*NEB*) whereas *CCGNRs* are shown to possess positive exchange bias (*PEB*). In addition, the electron spin resonance (*ESR*) signal of *GNRs* deviates from Lorentzian shape, however, the *ESR* signal from *CCGNRs* is fitted well by the Lorentzian shape. We have attributed the origin of such behavior to the atoms present at the edges of the ribbons. In *GNRs*, the edges are terminated by hydrogen and in *CCGNRs*, the edges are terminated by oxygen. In this letter, we have extended our efforts to address the dynamical magnetic properties of GNRs, which has not been done before, as far as we know.

Exchange bias (*EB*) and training effect (*TE*) are fundamental magnetic coupling phenomenon, and they are usually observed in mixed magnetic metallic systems[23 - 26], in which *AFM* and *FM* phases coexist. The *TE* can be manifested as the reduction of the *EB* field ($H_{EB}$) upon progressive field cycling. From the variation of $H_{EB}$ *vs.* the number of field cycles (*n*), the transition from the non–equilibrium to equilibrium nature of the spin structure can be inferred. Although the *EB* and *TE* have been found in other magnetic-metallic systems[23 - 26], it has not been explored in *GNRs* where it is extremely important to understand the dynamics of edge spins as a function of magnetic field and temperature. That constitutes the goal of present work, which is to probe the dynamics of magnetic hysteresis in GNRs while unveiling the magnetic nature of *GNRs* so as to employ *GNRs* for anticipated spintronic or metamaterial applications. In our earlier work[20], *GNRs* have been shown to exhibit mixed magnetic phases, and this allowed us to investigate the dynamics of spins present at the edges of *GNRs*.

Salient features of the present work have been noted, namely, a pronounced *TE* is ascertained when cycling the *GNRs* through several sequential hysteresis loops and pertaining the field sweep rate (*r*) dependence, at low *r* values, training of the exchange bias is rather high; however, the *TE* is not spread over a large number of cycles at high *r* values. In the present work, we compare such dynamical properties in GNR with those of chemically converted graphene nanoribbons[20] (CCGNRs), and are found to be absent.

Concisely, the preparation of *GNRs* involves the sealed-tube heating of multi-walled carbon nanotubes (MWCNTs) (with outside diameter of 40 – 80 nm and 15 – 20 inner nanotube layers) together with potassium metal in a furnace at $250^o$C for 14 h, followed by quenching to affect the longitudinal splitting process. The splitting process was further assisted by the generation of $H_2$ upon the ethanolic quench. The split MWCNTs were further exfoliated to form *GNRs* upon sonication in chlorosulfonic acid. Atomic force microscopy (AFM) and scanning electron microscopy (SEM) images show that the *GNRs* have widths of 100 - 250 nm and a length of 1-5 μm. *GNRs* were characterized with various techniques to test their electronic properties, as reported elsewhere[21]. *CCGNRs* were prepared by longitudinal unzipping of MWCNTs[22]. Briefly, this method involves the treatment of MWCNTs, consisting of 15-20 concentric cylinders and 40-80 nm diameter, with concentrated $H_2SO_4$ and $H_3PO_4$ followed by oxidation with $KMnO_4$ , and subsequently reduced by $N_2H_4$, to afford the *CCGNRs*.

A vibrating sample magnetometer (VSM) was used to measure the magnetization *vs*. magnetic field (M *vs* H) at 5 K and in the range of -1 to 1 T. The field sweep rate (*r*) was varied in the range of 0.1-0.7 T/min. Before each run, the sample was warmed to 300 K and then cooled down to desired temperature in order to avoid remnant effects.

To better understand *EB* and *TE* of *GNRs* and *CCGNRs*, further extensive magnetization measurements have been performed. The *TE* of an *EB* system is due to the non-equilibrium nature of spin structure which exists at the pinning layer and this can be manifested as the gradual decrease in the exchange bias field ($H_{EB}$) upon repeated progressive field cycling[23 - 26]. With the field cycling, there would be a change in the state of the pinning layer from non–equilibrium initial state to the quasi-equilibrium state through intermediate states. The change in the $H_{EB}$ is predominantly high between the 1st and 2nd loops; however, this change would be minor for higher loops. The minor change in $H_{EB}$ for higher loops would follow the power-law behavior[27] as a result of rearrangements of spin structure upon consecutive field cycling, causing fluctuations in the *FM–AFM* coupling.

The dynamic non–equilibrium properties of *GNRs* and *CCGNRs* are investigated via sweep rate (*r*) dependence of $H_{EB}$. To unveil the *TE,* initially, we cooled the sample from 300 to 5 K in the magnetic field of 1 T at various other field sweep rates 0.1–0.7 T/min. The sample was placed in a zero–field (ZF) environment for at least 20 h so that the magnetic state of the sample would come to the initial state before the next set of measurements. Fig. 1 presents the variation of magnetization (M) as a function of magnetic field (H) recorded at temperature of 5 K, with 0.1 – 0.7 T/min, collected for ten sequential loops (*n* = 10). The aim of this particular experiment is to track the variation in $H_{EB}$ as a function of *n,* a commonly observed phenomenon in mixed magnetic systems[23 – 26]. As depicted in the zoomed version shown in the inset of Fig.1, the hysteresis loop shift along the negative magnetic field axis is evident. However, such shift is not detected along the positive field axis, as observed for CCGNRs reported[20] in our earlier work. The shift (EB) is found to decay upon sequential field cycling (*n*). More importantly, this shift is found to occur at different *r* values as well, and the results pertaining to the field sweep rate dependence will be discussed below. To our surprise, such pronounced *TE* is not observed in the case of *CCGNRs* though intensively sought, and hence, will not be discussed further. The reason for the apparent absence of such *TE* in *CCGNRs* is unclear, though the *CCGNRs* have far more basal plane disruptions along with oxidized edges, as compared to the pristine basal planes and hydrogen atom-terminated edges of reductively prepared *GNRs*.

The value of $H_{EB}$ at each *n* is calculated using the formula $H_{EB} = (H^1_C – H^2_C)/2$, where $H^1_C$ and $H^2_C$ are the left and right coercive fields of the hysteresis loop. Fig. 2 shows the variation of $H_{EB}$ *vs. n*, collected at various *r* values in the range of 0.1–0.7 T/min and at 5 K. As it can be seen, this variation is not uniform for all the *n* values. At this point, we may separate the variation in $H_{EB}(n)$ into two regimes. In the first regime (up to *n* = 2), the decrease in $H_{EB}$ is higher, however, in the second regime (n>2) the decrease in $H_{EB}(n)$ is only minor.

In order to get further insights, and to quantify the results, we used a power-law behavior[28] to fit the TE behavior for *n* > 1 at various *r* values in the range 0.1–0.7 T/min. Empirically, the *TE* can be quantified by a power law function[28] $H_{EB}(n) = H_{EB}{}^\infty + Dn^{-\alpha}$, where $H_{EB}{}^\infty$ is the limiting value of $H_{EB}$, when the number of cycles *n* approaches infinity, and $\alpha$ is a positive exponent whose best fitting value is about 0.5. As depicted in Fig. 2, the solid red line shows the best fitting result for *n* >1. The fit shows a very good agreement with the experimental data. The inferred values of $H_{EB}{}^\infty$ are 1.12, 1.11, 1.06, 1.28 and 1.283 mT at different *r* values of 0.1, 0.3, 0.5, 0.6 and 0.7 T/min, respectively, with α value of 0.5.

As it can be noticed from the Fig. 2, there exists a steep variation in the $H_{EB}$ in the first regime and this cannot be explained by the power-law behavior alone. To account for such a steep relaxation in the *FM/AFM* exchanged coupled systems, Binek and co-authors have proposed[23, 24] a recursive equation that describes the dependence of $H_{EB}$ on *n*, also called the 'spin configurational relaxation model' (SCRM), as given below,

$$H_E(n+1) = H_E(n) - \gamma[H_E(n) - H_{EB}^\infty]^3 \qquad (1)$$

Where *γ* is a sample dependent constant. One can re-write the equation (1) as

$$\gamma = \frac{H_{EB}(n) - H_{EB}(n+1)}{[H_{EB}(n) - H_{EB}^\infty]^3} \qquad (2)$$

Using equation (2), a $\gamma$ value is extracted at various $r$ values for GNRs. A theoretical value of $H_{EB}$ is calculated by substituting the $\gamma$ and $H_{EB}{}^{\infty}$ in equation (1). The calculated data (solid red circle) exactly matches the experimental data (open circles) not only for $n > 1$ but also for $n = 1$ in the entire sweep range that was investigated. This shows that the *TE* in *GNRs* could be satisfactorily described by SCRM. From the data gathered, we may infer that the physical phenomenon of *TE* in our system could be due to the non-equilibrium nature of the spins in GNRs, similar to other systems[23,24] reported thus far. Upon sequential hysteresis loop cycling, a decrease in the $H_{EB}$ with $n$ is evident, and it could be attributed to the rearrangements in the spin structure of the *GNRs* towards equilibrium configuration. The SCRM is found to be applicable at all other sweep rates, as demonstrated in Fig.2.

Now we turn our attention to the discussion of the observed prominent *TE* at various field sweep rates. We explain this behavior by the dimensionless parameter $\gamma$ extracted from SCRM at various $r$ values. From equation (2), a high value of $\gamma$ requires small values for the denominator; this means that the deviation from the equilibrium state is less upon consecutive field cycling. On the other hand, it hints that *TE* is weaker if the value of $\gamma$ is high. In contrast to that, lower value of $\gamma$ would afford a higher value for the denominator in equation (2), strong training effects would be evident.

Fig. 3a shows the variation of $\gamma$ with respect to $r$. For the low $r$ values, a lower value of $\gamma$ is evident. This indicates that *TE* is significant. Contrary to that, $\gamma$ value decreases with $r$ up to 0.5 T/min and above this value $\gamma$ increases. This can be interpreted in such a way that the related *TE* is stronger at low $r$ values, and above 0.5 T/min the *TE* is weaker. At low $r$ values, the absolute *TE* is found to be large, which is, however, spread over a large $n$. Nevertheless, for higher $r$ values and above 0.5 T/min, the $H_{EB}$ value is constant after certain $n$ values. To provide further information, the steepness parameter[28] can be defined as $C = (H_{EB} (n = 1) - H_{EB} (n = 2))/(H_{EB} (n = 1) - H_{EB}^{\infty})$. If the value of $C = 1$, which means that there would be a step-like change between the first two values of $H_{EB}$ upon consecutive field cycling. Nevertheless, if $C < 1$, a gradual decrease in the $H_{EB}$ can be obtained. In our case, the value of $C$ (~ 0.3) is less than 1 for all the $r$ values, indicating that the gradual change in $H_{EB}$ upon repeated field cycling is noticed.

In the present letter, the exchange bias training effect of *GNRs* has been explained by Landau – Khalatnikov (LK) theory using power-law dependence in order to characterize the time evolution of interface magnetization in the anti-ferromagnetic layer when *GNRs* approach equilibrium, as it has been well-established by Binek[23] *et al.* and Xi[29] *et al.* Basically, the spins at the ribbon edges align ferromagnetically (*FM*) or antiferromagnetically (*AFM*) and such configuration of the spins may lead to an interaction between two ordered states (*FM* or *AFM*). This would indeed results in the pinning of the *FM* spins at *FM/AFM* interface regions, such regions are responsible for the exchange bias phenomenon in *GNRs*, and they lead to training behavior upon repeated field cycling as a consequence of rearrangements in the spin structure of the *GNRs* toward equilibrium configuration. Xi *et al.* have provided an alternative explanation while studying the training effect in NiFe/IrMn bilayers. These are based on the nucleation and domain dynamics of AF grains as suggested by the Kolmogorov – Avrami (KA) model[30]. This model appears to be much more refined with better theoretical understanding incorporated, questioning the simple power-law dependence. However, in the current work, we did not attempt to test this KA model, and that forms the subject of our near future work on *GNRs*. Both the *GNRs* and NiFe/IrMn systems are entirely different, particularly the spin-orbit coupling of NiFe/IrMn is much higher than that of GNRs, which is directly related to the magnetic anisotropy and magnetic coerceivity; though both are polycrystalline materials in nature. This essential difference may have direct influence on the unidirectional exchange anisotropy (exchange bias) and its dynamical property, i.e. training effect and relaxation of *FM/AFM* magnetization.

Sweep rate dependence of the exchange bias has been studied in several other exchange coupled systems[29 - 32]. At each and every sweep rate, the measured value of $H_{EB}$ is the averaged value for last nine sequential loops and plotted as a function of $r$ in Fig. 3b. Variation of the $H_{EB}$ with $r$ could be satisfactorily explained by the following power–law equation[33], given below,

$$H_{EB} = H_{EB}^0 r^\beta \qquad (3)$$

where $H_{EB}^0$ is the limiting value of the exchange bias, $r$ is field sweep rate and $\beta$ is a constant, respectively. Incidentally, we could well-explain our results with the above power–law equation. As shown in Fig. 3b, the open circles are the experimental data and red solid curve indicates the best least square fitting, experiment and theory are found to be in a good agreement with each other. The obtained values of the $H_{EB}^0$ and $\beta$ are 2.51 mT and 0.051 respectively. A similar kind of power–law dependence of $H_{EB}$ with the field sweep rate has been observed in $Ni_{81}Fe_{19}/Ir_{22}Mn_{78}$ bilayers[29]. According to Xi and co-authors[29], the observed change in the $H_{EB}$ with $r$ can be related to the relaxation time of the anti-ferromagnetically aligned spins in GNRs. Furthermore, Mc-Michael *et al.* have recognized[34] that the difference in time that it takes to perform a hysteresis loop measurement compared with the relaxation time of the anti-ferromagnetically aligned spins can lead to a change in the $H_{EB}$ with the applied field sweep rate. From our experimental evidence described above, we believe that a similar mechanism is operative in the present case as well.

In summary, the dynamical magnetic properties such as the training effect and magnetic field sweep rate dependence of exchange bias of exchange coupled graphene nanoribbons are investigated. The obtained results are well-explained by the well-known spin configurational relaxation model. The training effect is more pronounced for the low field sweep rate; however, the training effect is not prominent over more number of cycles at high field sweep rate. The increase in the exchange bias field with the field sweep rate obeyed the power–law behavior. The present results pertinent to the dynamical response of the exchange bias in GNRs are important for broadening our current understanding of the magnetism in graphene nanoribbons and may pave the way for possible device applications, upon appropriately engineering the edge magnetism and edge spin dynamics.

SNJ would like to thank KU Leuven, for research fellowship, This work is supported by the K.U. Leuven Excellence financing (INPAC), by the Flemish Methusalem financing and by the IAP network of the Belgian Government. JMT thanks Mitsui & Co., Ltd., M. Endo and A. Tanioka for generously donated the starting MWCNTs, the AFOSR (FA9550-09-1-0581) and the Office of Naval Research Graphene MURI Program (00006766) for support.

**Figure captions**

**Fig. 1:** Exchange bias training effect of the GNRs for ten sequential field cycles after field cooling the sample from 300 K in the presence of 1 T. The inset shows the systematic shift (in the direction of arrow) of the hysteresis loop along the magnetic field axis upon sequential field cycling.

**Fig. 2:** Variation of the exchange bias field ($H_{EB}$) as a function of progressive field cycling ($n$) at 5 K, collected at various field sweep rates. As depicted, the training effect (TE) is large at low field sweep rate and is limited for few cycles at high field sweep rate.

**Fig. 3:** (a) Variation of the dimensionless parameter γ as a function of field sweep rate. (b) Variation of the exchange bias field ($H_{EB}$) with the field sweep rate ($r$). The solid red curve is resulted from the fit of power–law behavior.

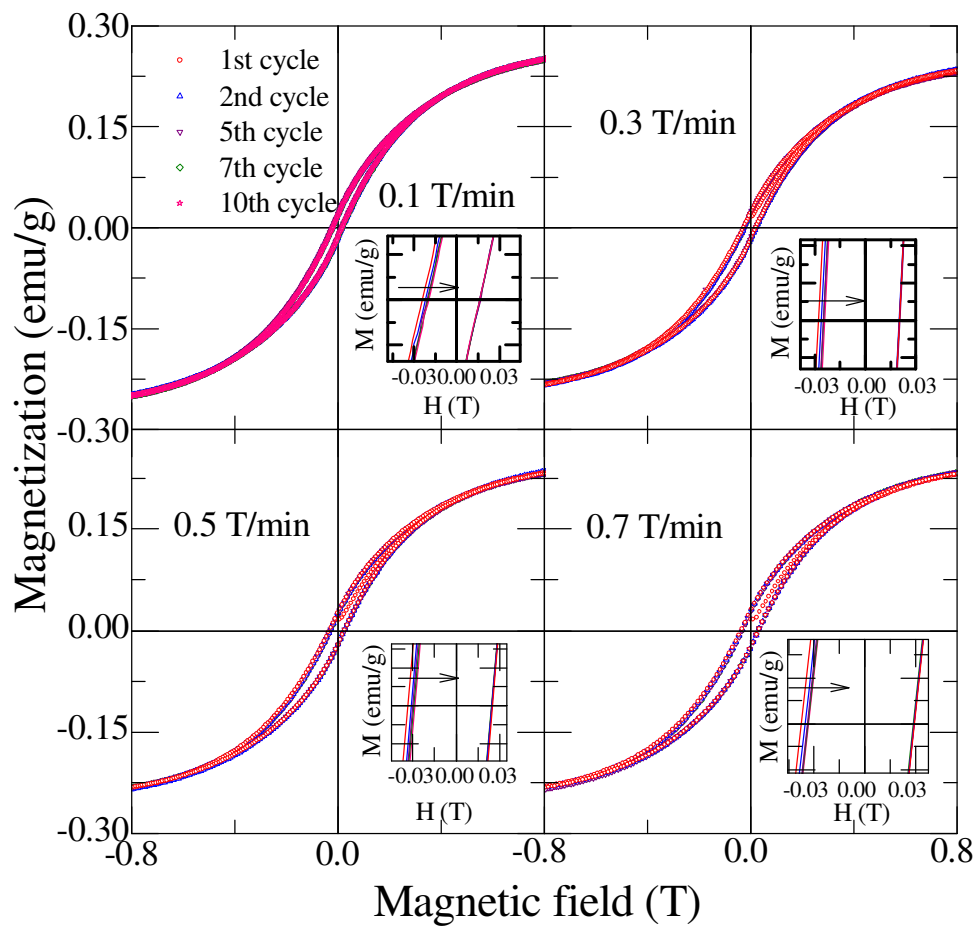

**Fig. 1:** Exchange bias training effect of the GNRs for ten sequential field cycles after field cooling the sample from 300 K in the presence of 1 T. The inset shows the systematic shift (in the direction of arrow) of the hysteresis loop along the magnetic field axis upon sequential field cycling.

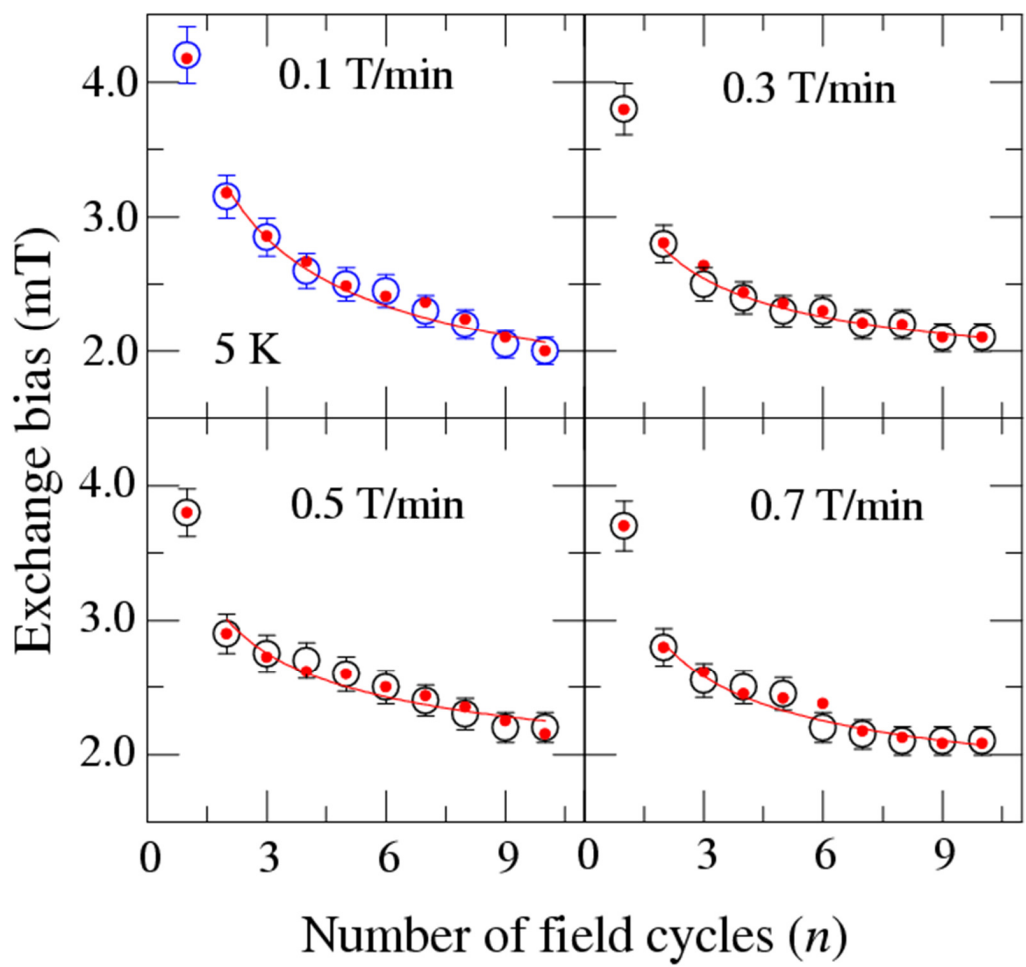

**Fig. 2:** Variation of the exchange bias field (H$_{EB}$) as a function of progressive field cycling (n) at 5 K, collected at various field sweep rates. As depicted, the training effect (TE) is large at low field sweep rate and is limited for few cycles at high field sweep rate.

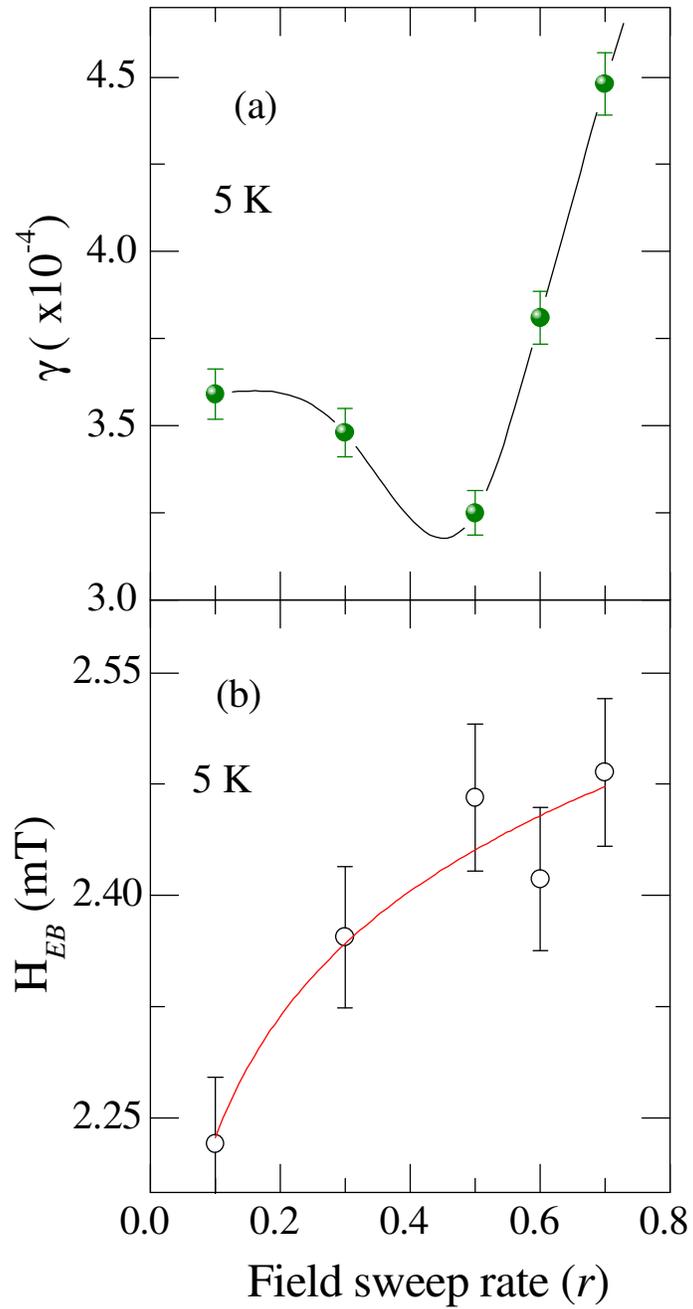

**Fig. 3:** (a) Variation of the dimension less parameter γ as a function of field sweep rate. (b) Variation of the exchange bias field (H$_{EB}$) with the field sweep rate ($r$). The solid red curve is resulted from the fit of power–law behavior.